\documentstyle[amssymb,aps]{revtex}

\draft

\begin{document}

\tighten
\twocolumn[\hsize\textwidth\columnwidth\hsize\csname@twocolumnfalse\endcsname
\title{Expansion of a quantum degenerate boson-fermion mixture}
\author{Hui Hu$^1$, Xia-Ji Liu$^{2,3}$ and Michele Modugno$^4$}
\address{ $^1$Abdus Salam International Center for Theoretical
Physics, P. O. Box 586, Trieste 34100, Italy\\ 
$^2$LENS, Universit\`{a} di Firenze, Via Nello Carrara 1, 50019 Sesto
Fiorentino, Italy\\ 
$^3$Institute of Theoretical Physics, Academia
Sinica, Beijing 100080, China\\
$^4$INFM, LENS and Dipartimento di Fisica,
Universit\`{a} di Firenze,\\ Via Nello Carrara 1, 50019 Sesto
Fiorentino, Italy } 
\date{\today} 
\maketitle

\begin{abstract}
We study the expansion of an ultracold boson-fermion mixture released from
an elongated magnetic trap, in the context of a recent experiment at LENS
(G. Roati {\it et al.}, Phys. Rev. Lett. {\bf 89}, 150403 (2002)). We
discuss in some details the role of the boson-fermion interaction on the
evolution of the radial-to-axial aspect ratio of the condensate, and show
that it depends crucially on the relative dynamics of the condensate and
degenerate Fermi gas in the radial direction, which is characterized by the
ratio between the trapping frequency for fermions and for bosons. Our
numerical simulations are in reasonable agreement with the experiment.
\end{abstract}

]

\section{Introduction}

The experimental realization of ultracold bose-fermi mixtures of alkali
atoms introduces a novel tool for the study of various quantum phenomena 
\cite{hulet,ketterle,roati,modugno}. The most appealing one is certainly the
Bardeen-Cooper-Schrieffer (BCS) type superfluidity, since the presence of
boson-fermion interaction in the mixture is expected to induce an effective
attraction between fermions by exchanging density fluctuations of the
bosonic background \cite{heiselberg,bijlsma}, analogously to that of
superfluid $^3${\rm He} and $^4${\rm He} in low-temperature physics \cite
{edwards,anderson}. With this respect, the recently reported quantum
degenerate mixtures of $^{40}${\rm K} (fermion) and $^{87}${\rm Rb} (boson)
by the LENS group would be particularly interesting due to the {\em large}
boson-fermion attraction, which has been directly manifested by the collapse
of the degenerate fermions above a critical number of particles \cite
{modugno}, or alternatively, by the observation that the Bose-Einstein
condensate (BEC) coexisting with the Fermi gas inverts its aspect ratio
more rapidly than a pure condensate during the ballistic expansion \cite
{roati}. In the latter case, the experimental result agrees qualitatively
with the expectations of a tighter confinement for a BEC in a Fermi gas with
mutual attraction \cite{roth,roth2}, and is qualitatively fitted by the
theoretical predictions for the free expansion of a pure BEC with trap
frequencies $10\%$ larger than the actual one. However, possible effects of
the boson-fermion attraction {\em during} the early stages of the expansion
are not considered in the fitting and certainly need a further investigation.

In this paper, we would like to study in details the effect of the
boson-fermion attraction on the expansion of the condensate as well as of
the Fermi gas after the release from the strongly elongated magnetic trap.
We have taken into account both ground state effects and the effect of
attraction between bosons and fermions during the early stages of the
expansion. The latter one is found to play an important role on the
evolution of the radial-to-axial aspect ratio of the condensate. Our
numerical simulations are in reasonable agreement with the experiment.

\section{Formulation}

We consider a dilute spin-polarized boson-fermion binary mixture at very low
temperature trapped in a strongly elongated harmonic oscillator potential.
In the semi-classical Thomas-Fermi approximation, the condensate and
degenerate Fermi gas evolve according to the Stringari's hydrodynamic
formulation \cite{stringari} 
\begin{eqnarray}
&&\frac{\partial n_b}{\partial t}+\nabla \cdot \left( n_b{\bf v}_b\right)
=-\Gamma _{coll},  \nonumber \\
&&m_b\frac{\partial {\bf v}_b}{\partial t}+{\bf \nabla }\left(
V_{ho}^b+g_{bb}n_b+g_{bf}n_f+\frac 12m_b{\bf v}_b^2\right) =0,
\label{stringari}
\end{eqnarray}
and Boltzmann-Vlasov kinetic equation \cite{odelin,menotti}, 
\begin{equation}
\frac{\partial f}{\partial t}+{\bf v}_f\cdot \frac{\partial f}{\partial {\bf %
r}}-\frac 1{m_f}\frac{\partial V_{ho}^f}{\partial {\bf r}}\cdot \frac{%
\partial f}{\partial {\bf v}_f}-\frac{g_{bf}}{m_f}\frac{\partial n_b}{%
\partial {\bf r}}\cdot \frac{\partial f}{\partial {\bf v}_f}=I_{coll},
\label{boltzmann}
\end{equation}
respectively. Here $f\left( {\bf r},{\bf v}_f,t\right) $ is the single
particle phase space distribution function for fermions; $n_b\left( {\bf %
r},t\right) $, $n_f\left( {\bf r},t\right) =\int d^3{\bf v}_ff\left( {\bf r},%
{\bf v}_f,t\right) $ and ${\bf v}_b\left( {\bf r},t\right) $, ${\bf v}%
_f\left( {\bf r},t\right) $ are the densities and velocity
fields for bosons and fermions, respectively. 
Here, for simplicity, we consider the case in which they are trapped
by symmetric confining potentials,
$V_{ho}^{b,f}\left( {\bf r}\right) =\frac 12%
m_{b,f}\left( \omega _{\bot b,f}^2\rho ^2+\omega _{zb,f}^2z^2\right) $,
in a {\em concentric} configuration. Note that in the realistic
experimental situation the centers of mass of the condensate and Fermi gas
are instead {\em displaced} due to the different gravitational sag 
for $^{40}${\rm K} and $^{87}${\rm Rb} \cite{roati}. 
The displacement is not sufficiently large to affect the geometrical 
overlap of the two degenerate species. However, in
some degree it reduces the boson-fermion attraction. Of course,
a quantitative analysis should take into account this displacement. We
will comment on this point later.

The influence of the interatomic scattering processes between the two
species on their dynamics has been taken into account in two parts: mean
fields and collisions. In the above equations, the term $g_{bf}n_f\left( 
{\bf r},t\right) $ and $g_{bf}n_b\left( {\bf r},t\right) $ are included as
the Hartree-Fock mean field effect of boson-fermion interaction \cite{molmer}
(note that the latter one is known as the Vlasov contribution in the
literatures). The boson-boson and boson-fermion interaction strength for the
pseudopotentials, $g_{bb}$ and $g_{bf}$, are related to the $s$-wave
scattering lengths $a_{bb}$ and $a_{bf}$ through $g_{bb}=4\pi \hbar
^2a_{bb}/m_b$, $g_{bf}=2\pi \hbar ^2a_{bf}/m_{bf}$, in which $%
m_{bf}=m_bm_f/\left( m_b+m_f\right) $ is the reduced boson-fermion mass. On
the other hand, the dissipation term $-\Gamma _{coll}$ or $I_{coll}$
accounts for collisions \cite{note}. The relative importance of the mean
field and dissipation terms in describing the dynamics of whole system
depends upon the rate of collisions compared to the characteristic time
scale of the single particle oscillation. When collisions are relatively
rare, the mean-field interaction dominates and the system is in the
collisionless regime. In the opposite, hydrodynamic regime, the degenerate
Fermi gas is always in the local equilibrium with the condensate \cite{note2}%
. The dynamics is then most appropriately described by the hydrodynamic
Euler equation of motion \cite{amoruso,kagan}. Due to the lack of
information on the collision rate in experiment, we shall
mainly focus on the collisionless regime by neglecting the dissipation terms $%
-\Gamma _{coll}$ and $I_{coll}$. The opposite hydrodynamic regime will be
briefly discussed at the end of this paper.

In the absence of the boson-fermion interaction ($g_{bf}=0$) and dissipation
terms, both the Eqs. (\ref{stringari}) and (\ref{boltzmann}) admit a simple
scaling solution, {\it i.e.}, 
\begin{eqnarray}
n_b\left( {\bf r},t\right) &=&\frac 1{\prod_jb_j\left( t\right) }n_b^0\left( 
\frac{r_i}{b_i(t)}\right) ,  \nonumber \\
v_{bi}\left( {\bf r},t\right) &=&\frac 1{b_i(t)}\frac{db_i(t)}{dt}r_i,
\label{bansatz}
\end{eqnarray}
for the condensate \cite{kagan,castin} and 
\begin{eqnarray}
f\left( {\bf r},{\bf v}_f,t\right) &=&f_0\left( \frac{r_i}{\gamma _i(t)},%
{\bf V}({\bf r},t)\right) ,  \nonumber \\
V_i({\bf r},t) &=&\gamma _i(t)v_{fi}-\frac{d\gamma _i(t)}{dt}r_i,
\label{fansatz}
\end{eqnarray}
for the degenerate Fermi gas \cite{odelin,menotti,bruun}. Here $n_b^0$ and $%
f_0$ are the equilibrium distributions. The dependence on time $t$ is
entirely contained the six dimensionless scaling parameters, $b_i\left(
t\right) $ and $\gamma _i(t)$, where $i=x$, $y$, and $z$. By substituting
this solution into Eqs. (\ref{stringari}) and (\ref{boltzmann}), it is
easily to show that, the scaling parameters obey the differential equations 
\cite{castin,bruun}, 
\begin{eqnarray}
\ddot{b}_i(t)+\omega _{ib}^2(t)b_i(t)-\frac{\omega _{ib}^2(0)}{%
b_i(t)\prod_jb_j\left( t\right) } &=&0,  \label{bfree} \\
\ddot{\gamma}_i(t)+\omega _{if}^2(t)\gamma _i(t)-\frac{\omega _{if}^2(0)}{%
\gamma _i^3(t)} &=&0.  \label{ffree}
\end{eqnarray}
Solutions of Eqs. (\ref{bfree}) and (\ref{ffree}), respectively, determine
the time evolution of the pure condensate and Fermi gas. In particular, they
can be used to study the expansion of the system after a sudden and total
opening of the trap at $t=0$. For an elongated cylindrical trap with
anisotropic parameter $\lambda =\omega _{zb}/\omega _{\perp b}\left( =\omega
_{zf}/\omega _{\perp f}\right) \ll 1$, one may find that \cite{castin,bruun} 
\begin{eqnarray}
&&b_{\perp }(\tau ) =\sqrt{1+\tau ^2},  \nonumber \\
&&b_z(t) =1+\lambda ^2\left[ \tau \arctan \tau -\ln \sqrt{1+\tau ^2}\right] ,
\label{sb}
\end{eqnarray}
and 
\begin{eqnarray}
\gamma _{\perp }(\tau ) &=&\sqrt{1+\beta ^2\tau ^2},  \nonumber \\
\gamma _z(\tau ) &=&\sqrt{1+\lambda ^2\beta ^2\tau ^2},  \label{sf}
\end{eqnarray}
where we have introduced a dimensionless time variable $\tau =\omega _{\perp
b}t$ and $\beta =\omega _{\perp f}/\omega _{\perp b}$.

In the presence of the boson-fermion interaction ($g_{bf}\neq 0$), however,
the simple scaling solution is no longer satisfied at every position ${\bf r}
$ after the substitution. A useful approximation, in the first order of $%
g_{bf}$, is to assume the scaling form of the solution as a priori, and
fulfill it {\em on average} by integrating over the spatial coordinates.
This strategy has been recently used by Guery-Odelin \cite{odelin} to
investigate the effect of the interaction on the collective oscillation of a
classical gas in the collisionless regime and by Menotti {\it et al.} to
study the expansion of an interacting Fermi gas \cite{menotti}. In some
sense, this approximation is equivalent to the sum-rule approach \cite
{menotti} that is extensively used in evaluating the low-energy collective
modes of dilute quantum gases \cite{stringari}. We have also recently
applied this approximation to derive the coupled set of differential
equations for $b_i\left( t\right) $ and $\gamma _i(t)$, and studied the
monopole and quadrupole excitations of bose-fermi mixtures after linearizing
these equations around the equilibrium points \cite{liu}. Here we only
present the brief derivation to make the paper self-contained.

As specified above, we substitute the scaling ansatz Eq. (\ref{bansatz})
into Stringari's hydrodynamic equations. By setting $R_i=r_i/b_i(t)$, one
finds, 
\begin{eqnarray}
&&\ddot{b}_i(t)R_i+\omega _{ib}^2(t)b_i(t)R_i+\frac{g_{bb}}{m_b}\frac 1{%
b_i(t)\prod_jb_j\left( t\right) }\frac{\partial n_b^0\left( {\bf R}\right) }{%
\partial R_i}  \nonumber \\
&&\left. \hspace{0.8in}+\frac{g_{bf}}{m_b}\frac 1{b_i(t)\prod_j\gamma
_j\left( t\right) }\frac{\partial n_f^0(\frac{b_i\left( t\right) }{\gamma
_i\left( t\right) }R_i)}{\partial R_i}=0\right. .  \label{btmp}
\end{eqnarray}
The coupled differential equations for the scaling parameters $b_i(t)$ can
be obtained by multiplying Eq. (\ref{btmp}) by $R_in_b^0\left( R_i\right) $
on both sides and integrating over the spatial coordinates. Making use of
the equilibrium properties of the density distribution in the ground state, 
\begin{equation}
\omega _{ib}^2(t)R_i+\frac{g_{bb}}{m_b}\frac{\partial n_b^0\left( {\bf R}%
\right) }{\partial R_i}+\frac{g_{bf}}{m_b}\frac{\partial n_f^0({\bf R})}{%
\partial R_i}=0,  \label{btmp2}
\end{equation}
after some straightforward algebra one finds, 
\begin{eqnarray}
&&\ddot{b}_i(t)+\omega _{ib}^2(t)b_i(t)-\frac{\omega _{ib}^2(0)}{%
b_i(t)\prod_jb_j\left( t\right) }  \nonumber \\
&&+\frac{g_{bf}}{m_bN_b\left\langle R_i^2\right\rangle _b}\frac 1{%
b_i\prod_jb_j}\int d^3{\bf R}\frac{\partial n_f^0({\bf R})}{\partial R_i}%
R_in_b^0(\frac{\gamma _i}{b_i}R_i)  \nonumber \\
&&\left. -\frac{g_{bf}}{m_bN_b\left\langle R_i^2\right\rangle _b}\frac 1{%
b_i\prod_jb_j}\int d^3{\bf R}\frac{\partial n_f^0({\bf R})}{\partial R_i}%
R_in_b^0({\bf R})=0\right. ,  \label{bscaling}
\end{eqnarray}
where $\left\langle R_i^2\right\rangle _b=(1/{N_b})\int d^3{\bf R}n_b^0(%
{\bf R})R_i^2$ is the average size of bosons along the $i$-axis. The last
two terms in Eq. (\ref{bscaling}), linear in $g_{bf}$, account for the
effects of boson-fermion interaction.

Analogous procedure can also be applied to the fermionic part. We substitute
the scaling ansatz Eq. (\ref{fansatz}) into Boltzmann-Vlasov kinetic
equation to get 
\begin{eqnarray}
&&\frac{V_i}{\gamma _i^3(t)}\frac{\partial f_0}{\partial R_i}-\left[ \ddot{%
\gamma}_i(t)+\omega _{if}^2(t)\gamma _i(t)\right] R_i\frac{\partial f_0}{%
\partial V_i}  \nonumber \\
&&\left. -\frac{g_{bf}}{m_b}\frac 1{\gamma _i(t)\prod_jb_j\left( t\right) }%
\frac{\partial n_b^0(\frac{\gamma _i\left( t\right) }{b_i\left( t\right) }%
R_i)}{\partial R_i}\frac{\partial f_0}{\partial V_i}=0\right. ,  \label{ftmp}
\end{eqnarray}
where $R_i=r_i/b_i(t)$. Using the equilibrium properties of the distribution
function, 
\begin{equation}
V_i\frac{\partial f_0}{\partial R_i}-\omega _{if}^2(0)R_i\frac{\partial f_0}{%
\partial V_i}-\frac{g_{bf}}{m_b}\frac{\partial n_b^0(R_i)}{\partial R_i}%
\frac{\partial f_0}{\partial V_i}=0,  \label{ftmp2}
\end{equation}
to replace the first term of Eq. (\ref{ftmp}) by a linear superposition of $%
R_i({\partial f_0}/{\partial V_i})$ and $({\partial n_b^0(R_i)}/{%
\partial R_i})({\partial f_0}/{\partial V_i})$, and taking the moment of $%
R_iV_i$ (namely, $(1/{N_f})\int R_iV_i[Eq.(\ref{ftmp})]d^3{\bf R}d^3{\bf V%
}$), we finally obtain: 
\begin{eqnarray}
&&\ddot{\gamma}_i(t)+\omega _{if}^2(t)\gamma _i(t)-\frac{\omega _{if}^2(0)}{%
\gamma _i^3(t)}  \nonumber \\
&&+\frac{g_{bf}}{m_fN_f\left\langle R_i^2\right\rangle _f}\frac 1{\gamma
_i\prod_j\gamma _j}\int d^3{\bf R}\frac{\partial n_b^0({\bf R})}{\partial R_i%
}R_in_f^0(\frac{b_i}{\gamma _i}R_i)  \nonumber \\
&&\left. -\frac{g_{bf}}{m_fN_f\left\langle R_i^2\right\rangle _f}\frac 1{%
\gamma _i^3}\int d^3{\bf R}\frac{\partial n_b^0({\bf R})}{\partial R_i}%
R_in_f^0({\bf R})=0\right. ,  \label{fscaling}
\end{eqnarray}
where $\left\langle R_i^2\right\rangle _f=\frac 1{N_f}\int d^3{\bf R}n_f^0(%
{\bf R})R_i^2$.

The coupled set of differential equations (\ref{bscaling}) and (\ref
{fscaling}) is a generalization of Eqs. (\ref{bfree}) and (\ref{ffree}) in
the presence of the boson-fermion coupling. It determines the coupled
dynamics of the condensate and degenerate Fermi gas in the collisionless
regime as far as the assumption of the simple scaling solution is valid. We
shall only interested in the evolution of $b_i(t)$ and $\gamma _i(t)$ with
initial condition of $b_i(0)=1$, $\gamma _i(0)=1$, $\dot{b}_i(0)=0$, and $%
\dot{\gamma}_i(0)=0$ ($i=\perp $ or $z$), in case of switching off the trap
suddenly at $t=0$, {\it i.e.}, $\omega _{ib,f}(t>0)=0$. Our final aim is to
calculate the aspect ratio of the condensate and the Fermi gas defined as $%
\lambda {b_{\perp }(t)}/{b_z(t)}$ and $\lambda {\gamma _{\perp }(t)%
}/{\gamma _z(t)}$, which are actually measured in experiment. According to
Eqs. (\ref{bscaling}) and (\ref{fscaling}), the whole process of our
numerical calculations in this paper consists of three stages. First, one
has to find the equilibrium ground-state densities: $n_b^0\left( \rho
,z\right) $ and $n_f^0\left( \rho ,z\right) $ at very low temperature, which 
{\em approximately} satisfy the following coupled equations in the
Thomas-Fermi approximation \cite{molmer}, 
\begin{eqnarray}
V_{ho}^b\left( \rho ,z\right) +g_{bb}n_b^0\left( \rho ,z\right)
+g_{bf}n_f^0\left( \rho ,z\right) &=&\mu _b,  \nonumber \\
\frac{\hbar ^2}{2m_f}\left( 6\pi ^2n_f^0\left( \rho ,z\right) \right)
^{2/3}+V_{ho}^f\left( \rho ,z\right) +g_{bf}n_b^0\left( \rho ,z\right)
&=&\mu _f,  \label{gs}
\end{eqnarray}
where $\mu _{b,f}$ is the chemical potential. It is convenient to obtain the
solutions of Eq. (\ref{gs}) by iterative insertion of one density
distribution in the other equation and numerically searching for $\mu _b$
and $\mu _f$ yielding the desired number of particles. Then one traces the
evolution of $b_i(t)$ and $\gamma _i(t)$ from $t$ to $t+\Delta t$ by
evaluating the integrations in Eqs. (\ref{bscaling}) and (\ref{fscaling}),
which turns out to be the most time-consuming step in the calculations. At
the final stage, one computes the radial-to-axial aspect ratio.

\section{Result and discussion}

In this work, we have performed numerical simulations with the parameters
chosen to reproduce the experimental conditions \cite{roati}. The number of
bosons and fermions in experiment ($N_b=2\times 10^4$ and $N_f=10^4$) are
large enough to ensure the validity of the Thomas-Fermi approximation, {\it %
i.e.}, $N_ba_{bb}/a_{ho,\bot }^b\gg 1$ and $N_f\gg 1$ \cite
{molmer,nygaard,butts}. We take the harmonic oscillator length $a_{ho,\bot
}^b=\sqrt{\hbar /(m_b\omega _{\bot b})}$ and $\hbar \omega _{\bot b}$ as
units, and introduce the quantities $\alpha =m_f/m_b$, $\beta =$ $\omega
_{\bot f}/\omega _{\bot b}$, and $\lambda =\omega _{zb}/\omega _{\bot
b}=\omega _{zf}/\omega _{\bot f}$ to parameterize the different mass of the
two components and anisotropy of traps. The constraint $\alpha \beta ^2=1$
is always satisfied since both bosons and fermions experience the same
trapping potential. As in experiment, we have $\alpha =0.463$, $\beta =1.47$%
, and $\lambda =0.0757$. We have also taken $a_{bb}=+110a_0$ and $%
a_{bf}=-330a_0$, where $a_0=0.529$ ${\rm \AA }$ is the Bohr radius \cite
{roati}. The most recent measurement suggested a new $s$-wave scattering
length $a_{bf}=-410a_0$ for the mixture of $^{40}${\rm K} and $^{87}${\rm Rb}
\cite{modugno,ferlaino}. However, the {\em effective} magnitude of $a_{bf}$
could be indeed lower, considering the possible effect of the
gravitational sag \cite{roati} and exchange correlations beyond the
mean-field approximation \cite{albus}. We will return to this point at 
the end of the paper.

Before presenting the numerical result, it is instructive to briefly analyze
the influence of the boson-fermion interaction on the expansion for both
species. In the collisionless regime, the effect of the dominated mean-field
interaction is two-fold: ({\it i}) First of all, it influences the profile
of the density distribution in the equilibrium ground state. In case of
attractive interaction, as shown in Fig. \ref{fig1}, both the density of the
condensate and of the degenerate Fermi gas are remarkably enhanced within
the central overlap region. The condensate profile narrows and the central
density increases moderately. The effect on the Fermi gas is more
pronounced: within the overlap region the fermionic density exhibits a
high-density bump on top of the low-density background and the central
density is increased by a factor of larger than two \cite{roth2}. Without
considering the mutual attraction during the expansion this enhancement of
the density profile, which corresponds to a tighter confinement, will
definitively lead to a faster expansion for both 
species if one turns off the trap potential
suddenly. In this paper, we shall consider such kind of fasten-mechanics as
a {\em static effect} 
of the boson-fermion interaction on the expansion. ({\it ii}%
) On the other hand, the condensate and degenerate Fermi gas also interact
with each other during the expansion (especially during the early stages). 
As a result, with the attractive boson-fermion interaction both
species will feel a {\em running} confinement potential generated by the
other species and therefore reduce their expansion rate. This slow-down
mechanics is determined by the relative dynamics between the condensate and
the degenerate Fermi gas. The smaller the relative expansion velocity is,
the more attraction the two species experience. With this respect, it is
referred to as a {\em dynamical effect} 
of the boson-fermion interaction on the
expansion.

Fig. \ref{fig2} shows the radial-to-axial aspect ratio
of the condensate and of the degenerate Fermi gas as a function of the
dimensionless expansion time variable $\tau $. Let us firstly concentrate on
the condensate. The evolution of the aspect ratio with $a_{bf}=-330a_0$ (the
dashed line in Fig. \ref{fig2}a) agrees qualitatively with the experimental
result (the solid circles). This agreement is remarkable, considering there
is no adjustable parameter in the numerical calculations and the simplicity
of our model.

In order to better understand the role of static and dynamical effect of the
mean-field boson-fermion interaction on the expansion of the condensate in
some details, we expand the density distribution of the Fermi gas around the
center by using the fact that in experiment the Fermi gas distributes more
widely than the condensate due to the Fermi statistics, and rewrite Eq. (\ref
{bscaling}) into a more physically transparent form, 
\begin{equation}
\ddot{b}_i(t)-\frac{\omega _{ib}^2}{b_i(t)\prod_jb_j(t)}+\frac{A\omega
_{ib}^2b_i(t)}{\gamma _i^2(t)\prod_j\gamma _j(t)}
-\frac{A\omega _{ib}^2}{%
b_i(t)\prod_jb_j(t)}\approx 0,  \label{bappr}
\end{equation}
where 
\begin{equation}A\approx \frac{g_{bf}}{m_bN_b\left\langle R_i^2\right\rangle
_b\omega_{ib}^2}\int d^3{\bf R}\frac{\partial n_f^0({\bf R})}
{\partial R_i}R_in_b^0({\bf R}).
\end{equation} 
Mathematically, the dynamical and static effects are represented
by the last two terms in Eq. (\ref{bappr}), respectively. Moreover, recalling
the free expansion of a pure condensate as shown in Eqs. (\ref{bfree}) and (%
\ref{sb}), the quantity $\omega _{ib}(1+A)^{1/2}$ can in fact be interpreted
as an effective trapping frequency experienced by the condensate in ground
state. The value of $A$ can further be roughly extracted from the change of
the bosonic density distribution due to the boson-fermion attraction. The
estimated value of $A\approx 40\%$ and the corresponding increase of the
trapping frequency $\Delta \omega /\omega \approx 20\%$ \cite{note3} is two
times larger than that used in the fitting in Ref.\cite{roati} as we
mentioned earlier in the beginning of introduction. This discrepancy can be
easily resolved if we consider the dynamical effect of the attraction on the
expansion. Indeed, by inserting the scaling solution for elongated traps 
at first order in $\lambda$ in Eq. (\ref{bappr}),  $b_{\perp }(\tau )=
\sqrt{1+(1+\delta)^2\tau ^2}$ 
($\delta $ represents the {\em net} increase
of the trapping frequency), $\gamma _{\perp }(\tau )\approx \sqrt { 1+\beta
^2\tau ^2}$, $b_z(\tau )\approx 1$, 
and $\gamma _z(\tau )\approx 1$, one obtains 
\begin{equation}
\delta =\beta \left[ \frac{\left( \beta ^4+4A(1+A)\right) ^{1/2}-\beta ^2}{2A%
}\right] ^{1/2}-1. 
\end{equation}
By substituting $\beta =1.47$ and $A\approx 40\%$ into the above equation, one
finds $\delta =12\%$, in good agreement with the value of $10\%$ used in Ref.%
\cite{roati}. In other words, on the expansion of the $^{87}${\rm Rb}
condensate, the static effect of the mean-field attraction dominates over
the dynamical one. If measured in units of increase of the trapping
frequency, they contribute around$+20\%$ and $-8\%$, respectively.

As explicitly shown in the third term in the left side of the Eq. (\ref
{bappr}), the dynamical effect of the mean-field attraction on the expansion
of the condensate is closely related to the relative dynamics between the
condensate and the degenerate Fermi gas, and specifically, related to the
value of $\beta =\omega _f/\omega _b$. In Fig. \ref{fig3} the
evolution of the aspect ratio of the condensate and of the degenerate Fermi
gas is plotted against the dimensionless expansion time. For both species,
the aspect ratio decreases as one lowers the value of $\beta $, suggesting
that the dynamical effect of the mean-field attraction becomes more and more
important with decreasing $\beta $. Precisely at $\beta =1$, where $b_{\perp
}(\tau )=\gamma _{\perp }(\tau )$, the static and dynamical effects for the
condensate are compensated with each other. As a result, the aspect ratio is
almost the same as that of a pure condensate. Note that although the above
result for $\beta \leqslant 1$ is not realistic for the mixture of $^{40}$%
{\rm K} and $^{87}${\rm Rb} in the LENS experiment, it might be relevant to
the system of $^6${\rm Li}-$^7${\rm Li} and $^{40}${\rm K}-$^{41}${\rm K},
in which $\beta \approx 1$.

The evolution of aspect ratio of the degenerate Fermi gas, on the
other hand, is also determined by the competition between the static
and dynamical effect of the mean-field attraction. As shown in
Figs. \ref{fig2}b and \ref{fig3}b, contrary to the condensate, the
dynamical effect for the Fermi gas always dominates and the aspect
ratio is less than that of a pure Fermi gas.

We now turn to consider the dependence of the aspect ratio on the
strength of boson-fermion interaction. In the first experiments at LENS 
\cite{roati}, only the absolute magnitude of $a_{bf}$, and not its sign, 
was {\em directly} determined. It is therefore interesting to
study the expansion also in case of repulsive interaction.
In Fig. \ref{fig2} the aspect ratio for $a_{bf}=+330a_0$ (dotted line) 
is compared with the result for $a_{bf}=-330a_0$, showing that 
the behavior of the latter is closer to the experiment points. 
The fact that boson-fermion interaction
is indeed negative has been subsequently confirmed by the observation
of the mixture collapse \cite{modugno}.
In order to understand which is the general behavior in passing from 
negative to positive values of the interspecies scattering length,
in Fig. \ref{fig4} we show the aspect ratio of the condensate and of 
the degenerate Fermi gas as a
function of $g_{bf}/g_{bb}$ at a fixed expansion time. 
For the condensate, the aspect ratio has a
parabolic shape with the minimum located at $g_{bf}/g_{bb}=0.5$. The
possible reason is that in the region of $g_{bf}/g_{bb}<0$ and 
$g_{bf}/g_{bb}>1$, the condensate is always tightly confined by either the
attractive or strong repulsive boson-fermion interaction and then the aspect
ratio is increased by the dominant static effect of mean-field interaction.
In contrast, the aspect ratio of the Fermi gas increases monotonically with
increasing the value of $g_{bf}/g_{bb}$ up to $g_{bf}/g_{bb}\approx 2$. It
is interesting to note that in principle it is possible to tune $g_{bf}$ by
using Feshbach resonances in experiment.

Up to now, we have restricted the discussion to the collisionless regime. In
the opposite hydrodynamic regime, the collision terms dominate over the
mean-field terms \cite{note2}. 
For the condensate, to ease the analysis we shall still
neglect the term $\Gamma _{coll}$ in the right side of Eq. (\ref{stringari}%
), by assuming that the dynamics of the condensate is less affected by the
collisions with the Fermi gas since the condensate always keeps itself in
the hydrodynamic regime. For the degenerate Fermi gas, we resort to the
Euler equation of motion \cite{amoruso,kagan}, which can be deduced from
Boltzmann-Vlasov kinetic equation under the assumption of local equilibrium.
The detailed expression of the coupled set of differential equations for $%
b_i\left( t\right) $ and $\gamma _i(t)$ in this regime can be found in Ref. 
\cite{liu}. The predictions of these equations are reported in Fig. 
\ref{fig5}. 
The aspect ratio of the condensate is affected by the collisions
only in a minor way, which might be understood from the assumption of the
vanishing value of $\Gamma _{coll}$. For the degenerate Fermi gas, in
contrast, the aspect ratio changes remarkably from the collisionless regime
to the hydrodynamic one. This is consistent with the result in Ref. \cite
{menotti}, where the {\em same} Euler equation of motion has been used to
describe the expansion of a {\em superfluid} Fermi gas \cite{note4}. Note
that the expansion of a strongly interacting degenerate Fermi gas has been
recently demonstrated in Ref. \cite{thomas}, and has been explained
qualitatively by the theory of Menotti {\it et al. }\cite{menotti}. However,
a stringent test of distinguishing the system from the normal hydrodynamic
phase to the superfluid phase is lacking in the experiment \cite{ps}.

\section{Summary}

In conclusion, we have studied the effect of the boson-fermion interaction
on the expansion of a boson-fermion mixture, within a simple scaling ansatz.
We have considered in some details the interplay of the so-called static and
dynamic effect of the mean-field interaction on the expansion. The former is
caused by the modified density profiles in the equilibrium
ground state, while the latter refers to the interaction between the
condensate and Fermi gas during the first stage of the expansion. These two
effects are compensated with each other. Which one plays the most important
role depends on the detailed parameters of the system.

For the mixture of $^{40}${\rm K} and $^{87}${\rm Rb}, the static effect is
dominant for the expansion of the condensate and its aspect ratio is
inverted more rapidly than that of a pure condensate. This feature has been
observed in the experiment, and also well reproduced by our numerical
simulations. For the degenerate Fermi gas, on the other hand, the dynamical
effect becomes important. As a result, its aspect ratio is less than that of
a pure Fermi gas. This prediction needs further experimental investigation.

At the end of this paper, several remarks are in order: ({\it i}) The above
results are based on the assumption of the simple scaling solution. The
justification of using such scaling ansatz on the problem of collective
excitations has been discussed in Ref \cite{liu} by the authors. We have
shown that this approximation is equivalent to the sum-rule approach.
However, on the problem of the expansion, its validity deserves a further
study (especially for the degenerate Fermi gas), though our numerical
results for the condensate are in reasonable agreement with the experiment. (%
{\it ii}) In a recent preprint, the importance of the corrections due to
exchange-correlation energy on the ground state of mixture $^{40}${\rm K}
and $^{87}${\rm Rb} has been discussed \cite{albus}. Such corrections on the
expansion can also be investigated by adding the exchange correlations in
the local density approximation. ({\it iii}) For the mixture of $^{40}${\rm K%
} and $^{87}${\rm Rb}, the recent measurements give a more accurate value of
the $s$-wave interspecies scattering length: $a_{bf}=-410a_0$ \cite
{modugno,ferlaino}. In Fig. \ref{fig6}, we show the aspect ratio for
such large negative scattering length without considering the effect of
different gravitational sag for $^{40}${\rm K} and $^{87}${\rm Rb}. The
result with inclusion of the exchange-correlation term is also plotted by
the dotted line. For the condensate the calculated aspect ratio is larger
than the experimental result. This discrepancy is {\em partly} due to the
possible effect of the gravitational sag \cite{note4}. ({\it iv}) Finally, a
careful consideration of the role of collisions is needed.

\section{Acknowledgments}

We acknowledge stimulating discussions with G. Modugno. One of us (X.-J.
Liu) wishes to thank the Abdus Salam International Centre for Theoretical
Physics (ICTP) for their hospitality during the early stages of this work.
X.-J. Liu is supported by the CAS K.C.Wang Post-doctoral Research Award
Fund, the Chinese Post-doctoral Fund and the NSF-China.

\begin{center}
{\bf Figures Captions}
\end{center}

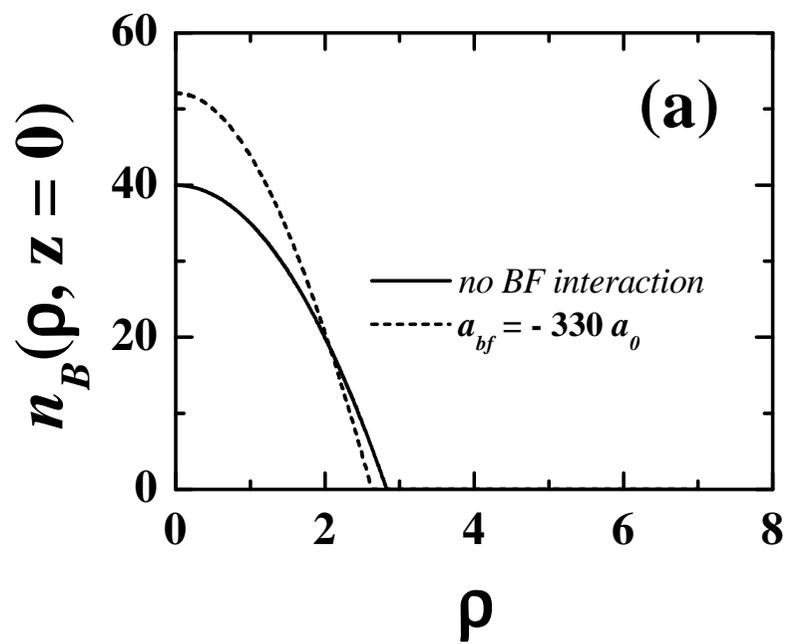
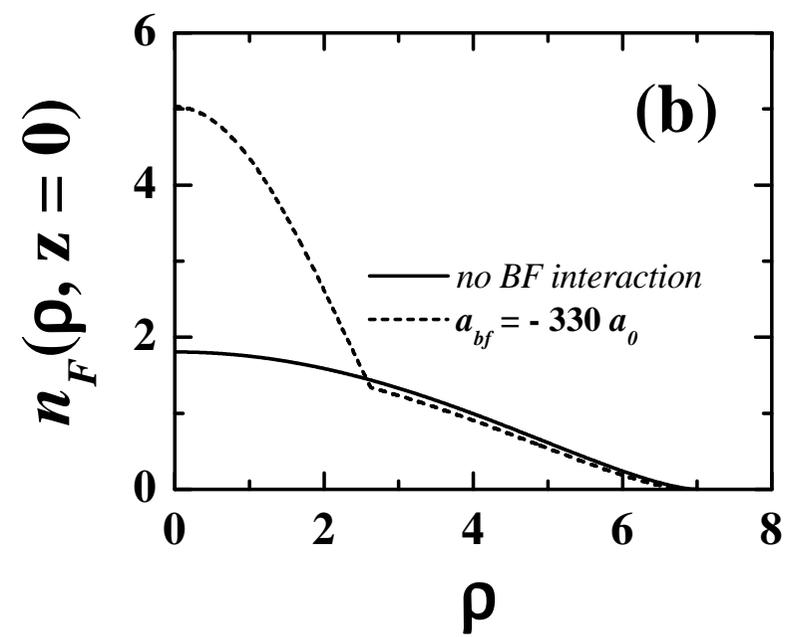
\begin{figure}
\caption{The equilibrium density distribution of the condensate (a) and of
the degenerate Fermi gas (b) along the radial direction for two different
values of $a_{bf}$: $a_{bf}=0$ (full line) and $a_{bf}=-330a_0$ (dashed
line), where $a_0=0.529$ ${\rm \AA }$ is the Bohr radius. We have taken $%
a_{bb}=110a_0=5.82$ {\rm nm}. The coordinate $\rho $ and the density $%
n_{b,f} $ are measured in units of the harmonic oscillator length $%
a_{ho,\bot }^b$ and $\left( a_{ho,\bot }^b\newline
\right) ^3$, respectively.}
\label{fig1}
\end{figure}

\begin{figure}
\caption{The radial-to-axial aspect ratio of the condensate (a) and of the
degenerate Fermi gas (b) as a function of the dimensionless expansion time
variable $\tau $ with a boson-fermion $s$-wave scattering length $a_{bf}=0$
(solid line), $-330a_0$ (dashed line), and $+330a_0$ (dotted line). For
comparison, the experimental data (solid circles) are also plotted. One unit
of $\tau $ corresponds to $0.738$ {\rm ms} in the experiment 
\protect\cite{roati}.
Note that the aspect ratio for the condensate with $a_{bf}=-330a_0$ agrees
qualitatively with the experiment, without any adjustable parameters. Note
also that in the experiment, the typical absorption image is taken at $\tau
=21$ $($or $t=15.5$ {\rm ms}$)$ for the condensate and at $\tau =6.1$ $($or $%
t=4.5$ {\rm ms}$)$ for the degenerate Fermi gas.}
\label{fig2}
\end{figure}

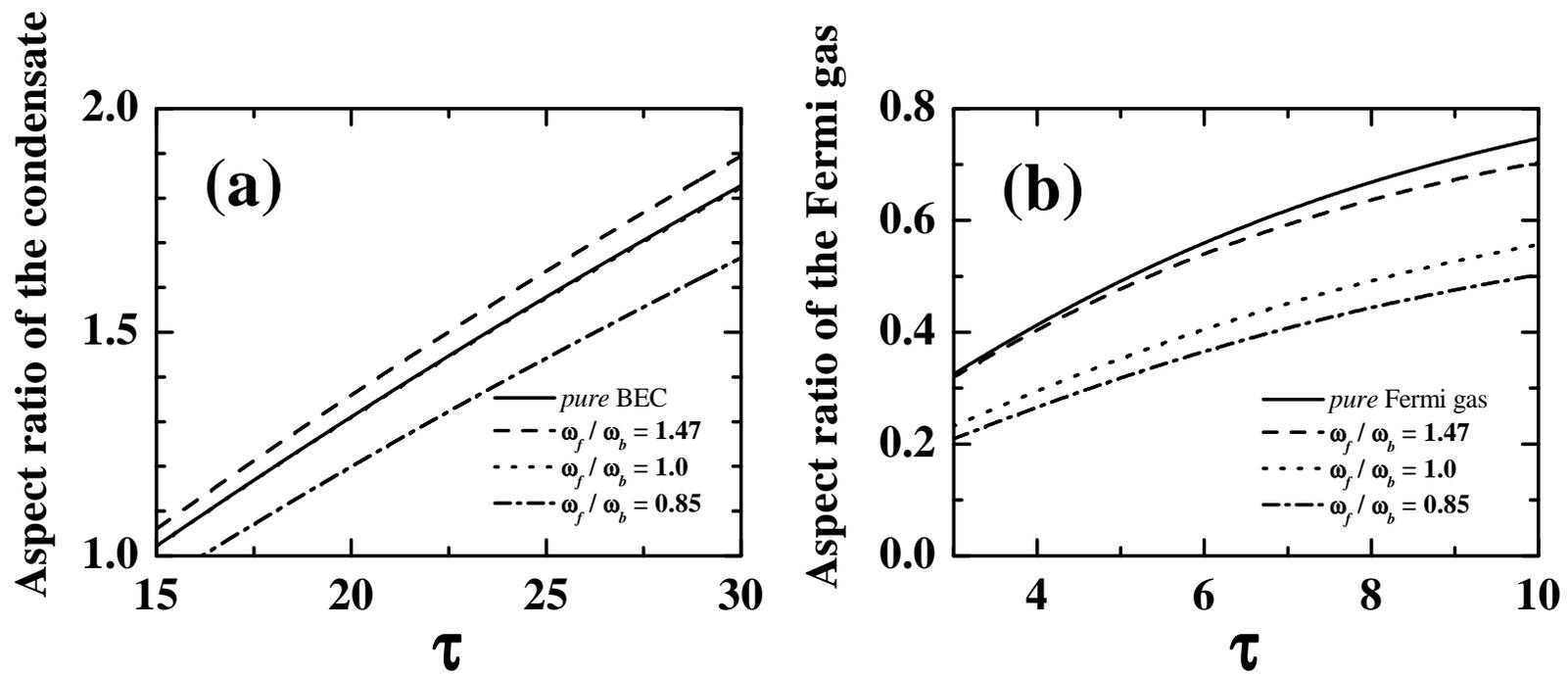
\begin{figure}
\caption{Aspect ratio of the condensate (a) and of the degenerate Fermi gas
(b) at $a_{bf}=-220a_0$ for various values of $\beta =\omega _{\perp
f}/\omega _{\perp b}$. The aspect ratio for $\beta =1.0$ in figure (a) is
almost the same as that of a pure BEC. Note that the value of $%
a_{bf}=-220a_0 $ is not realistic in experiment. However, if we take $%
a_{bf}=-330a_0$, we cannot find out the density distributions of the
equilibrium ground state for $\beta =1.0$ or $\beta =0.85$ within
Thomas-Fermi approximation. The possible reason is that the mixture composed
of $N_b=2\times 10^4$ and $N_f=10^4$ will collapse when $\beta \leqslant 1$.}
\label{fig3}
\end{figure}

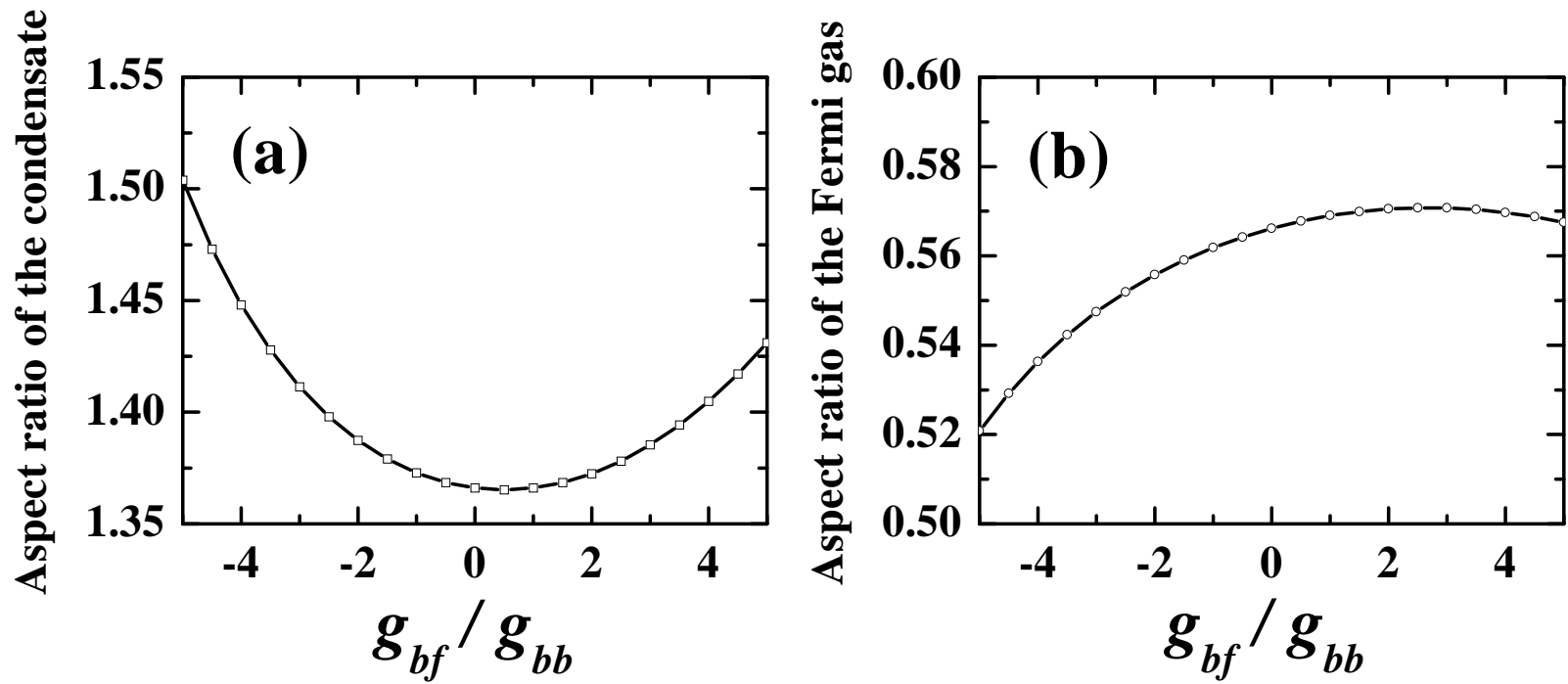
\begin{figure}
\caption{(a) The aspect ratio of the condensate at $\tau =21$ as a function
of $g_{bf}/g_{bb}$. (b) The aspect ratio of the Fermi gas at $\tau =6.1$ as
a function of $g_{bf}/g_{bb}$. In principle, it is possible to tune $g_{bf}$
by using Feshbach resonances. Note that $%
g_{bf}/g_{bb}=1.5875a_{bf}/a_{bb} $ for the mixture of $^{40}${\rm K} and $%
^{87}${\rm Rb}. In the experiment, $a_{bf}=-330a_0$ corresponds to $%
g_{bf}/g_{bb}=$ $-4.7625$.}
\label{fig4}
\end{figure}

\begin{figure}
\caption{Comparison of the aspect ratio of the condensate (a) and of the
degenerate Fermi gas (b) at $a_{bf}=-330a_0$ in the collisionless and
hydrodynamic regime. For comparison, the result for the decoupled
boson-fermion mixture in the collisionless regime are also plotted by the
solid lines.}
\label{fig5}
\end{figure}

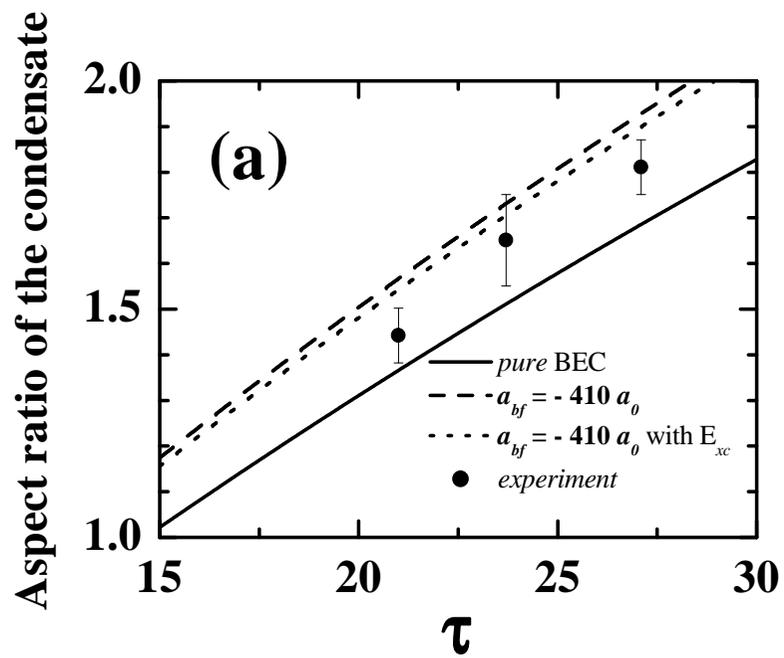
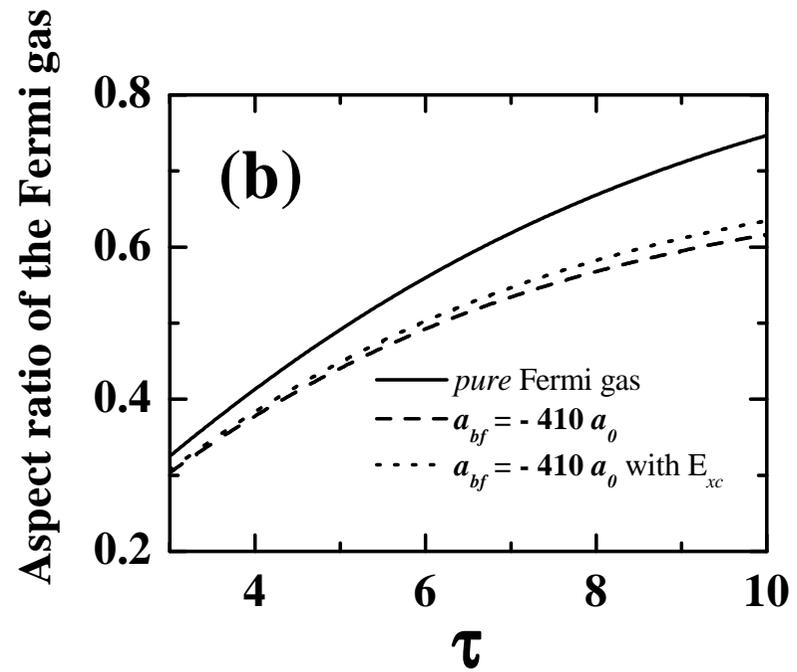
\begin{figure}
\caption{Aspect ratio of the condensate (a) and of the degenerate Fermi gas
(b) at $a_{bf}=-410a_0$ with and without the inclusion of the exchange
correlation term.}
\label{fig6}
\end{figure}


\begin{references}
\bibitem{hulet}  A. G. Truscott, K. E. Strecker, W. I. McAlexander, G. B.
Partridge, and R. G. Hulet, Science {\bf 291}, 2570 (2001); F. Schreck, L.
Khaykovich, K. L. Corwin, G. Ferrari, T. Bourdel, J. Cubizolles, and C.
Salomon, Phys. Rev. Lett. {\bf 87}, 080403 (2001).

\bibitem{ketterle}  Z. Hadzibabic, C. A. Stan, K. Dieckmann, S. Gupta, M. W.
Zwierlein, A. Gorlitz, and W. Ketterle, Phys. Rev. Lett. {\bf 88}, 160401
(2002).

\bibitem{roati}  G. Roati, F. Riboli, G. Modugno, and M. Inguscio, Phys.
Rev. Lett. {\bf 89}, 150403 (2002).

\bibitem{modugno}  G. Modugno, G. Roati, F. Riboli, F. Ferlaino, R. J.
Brecha, and M. Inguscio, Science {\bf 297}, 2240 (2002).

\bibitem{heiselberg}  H. Heiselberg, C. J. Pethick, H. Smith, and L.
Viverit, Phys. Rev. Lett. {\bf 85}, 2418 (2000).

\bibitem{bijlsma}  M. J. Bijlsma, B. A. Heringa, and H. T. C. Stoof, Phys.
Rev. A {\bf 61}, 053601 (2000).

\bibitem{edwards}  D. O. Edwards, D. F. Brewer, P. Seligman, M. Skertic, and
M. Yaqub, Phys. Rev. Lett. {\bf 15}, 773 (1965).

\bibitem{anderson}  A. C. Anderson, D. O. Edwards, W. R. Roach, R. E.
Sarwinski, and J. C. Wheatley, Phys. Rev. Lett. {\bf 17}, 367 (1966).

\bibitem{roth}  R. Roth and H. Feldmeier, Phys. Rev. A {\bf 65}, 021603(R)
(2002).

\bibitem{roth2}  R. Roth, Phys. Rev. A {\bf 66}, 013614 (2002).

\bibitem{stringari}  S. Stringari, Phys. Rev. Lett. {\bf 77}, 2360 (1996).

\bibitem{odelin}  D. Gu\'{e}ry-Odelin, Phys. Rev. A {\bf 66,} 033613 (2002).

\bibitem{menotti}  C. Menotti, P. Pedri and S. Stringari, Phys. Rev. Lett. 
{\bf 89}, 250402 (2002).

\bibitem{molmer}  K. M\o lmer, Phys. Rev. Lett. {\bf 80}, 1804 (1998).

\bibitem{note}  At low temperature, the main contribution to the collision
term comes from the collisions between a fermion and a condensed atom, in
which the fermion is scattered into another state, while the condensed boson
becomes a thermal atom. In the Hartree-Fock approximation, one may derive
the expressions for the dissipation terms as: $-\Gamma _{coll}=-\frac{%
n_cg_{bf}^2}{\left( 2\pi \right) ^5\hbar ^7}\int d{\bf q}d{\bf p}_1d{\bf p}%
_2\delta _\epsilon \delta _{{\bf p}}\times \left[ (1+\tilde{f}({\bf q}))(1-f(%
{\bf p}_2))f({\bf p}_1)-\tilde{f}({\bf q})(1-f({\bf p}_1))f({\bf p}%
_2)\right] $ and $I_{coll}=\frac{n_cg_{bf}^2}{\left( 2\pi \right) ^2\hbar ^4}%
\int d{\bf q}d{\bf p}_1d{\bf p}_2\delta _\epsilon \delta _{{\bf p}}\left(
\delta ({\bf p-p}_2)-\delta ({\bf p-p}_1)\right) \times \left[ (1+\tilde{f}(%
{\bf q}))(1-f({\bf p}_2))f({\bf p}_1)-\tilde{f}({\bf q})(1-f({\bf p}_1))f(%
{\bf p}_2)\right] ,$ where $\delta _\epsilon $ and $\delta _{{\bf p}}$ are
the usual delta functions accounting for the conservation of momentum and
energy. $\tilde{f}$ and $f$ are the single-particle phase-space distribution
function for thermal bosons and for fermions, respectively.

\bibitem{note2}  Here, the meaning of ``hydrodynamic'' refers only to the
Fermi gas. For the condensate, it is always in the hydrodynamic regime as
described by the Stringari's hydrodynamic equation.

\bibitem{amoruso}  M. Amoruso, A. Minguzzi, S. Stringari, M. P. Tosi, and L.
Vichi, Eur. Phys. J. D {\bf 4}, 261(1998).

\bibitem{kagan}  Yu. Kagan, E. L. Surkov, and G. V. Shlyapnikov, Phys. Rev.
A {\bf 55}, R18 (1997).

\bibitem{castin}  Y. Castin and R. Dum, Phys. Rev. Lett. {\bf 77}, 5315
(1996).

\bibitem{bruun}  G. M. Bruun and C. W. Clark, Phys. Rev. A {\bf 61}, 061601
(2000).

\bibitem{liu}  X.-J. Liu and H. Hu, cond-mat/0212169 (2002), to appear in
Phys. Rev. A.

\bibitem{nygaard}  N. Nygaard and K. M\o lmer, Phys. Rev. A {\bf 59}, 2974
(1999).

\bibitem{butts}  D. A. Butts and D. S. Rokhsar, Phys. Rev. A {\bf 55}, 4346
(1997).

\bibitem{ferlaino}  F. Ferlaino, R. J. Brecha, P. Hannaford, F. Riboli, G.
Roati, G. Modugno, and M. Inguscio, cond-mat/0211051 (2002).

\bibitem{albus}  A. P. Albus, F. Illuminati, and M. Wilkens,
cond-mat/0211060 (2002).

\bibitem{note3}  We have used two methods to extract the value of $\Delta
\omega /\omega $. One is to use the relation $n_b(0)\propto \omega ^{6/5}$,
and the other is $\left( \left\langle x^2+y^2\right\rangle \right)
^{1/2}\propto \omega ^{-2/5}$. Both of them are valid for the parabolic
confinement for a pure condensate. From Fig. \ref{fig1}a, the former relation
gives $\Delta \omega /\omega =25\%$, and the latter gives $\Delta \omega
/\omega =17\%$. The possible reason for the discrepancy might be due to the
fact that the density profile of the condensate in mixture is not perfectly
parabolic. Nevertheless, a rough estimate of $\Delta \omega /\omega \approx
20\%$ is convincing.

\bibitem{note4}  If the critical temperature for superfluidity is high
enough, both the normal the superfluid phases may be governed by the {\em %
same} hydrodynamic regime over the whole range of relevant temperatures.

\bibitem{thomas}  K. M. O'Hara, S. L. Hemmer, M. E. Gehm, S. R. Granade, and
J. E. Thomas, Science {\bf 298}, 2179 (2002).

\bibitem{ps}  L. Pitaevskii and S. Stringari, Science {\bf 298}, 2144 (2002).

\bibitem{note5}  Since it is hard to study the possible effect of the
gravitational sag on the expansion, we have considered the effect of
gravitational sag on the equilibrium density profile in the ground state
without the inclusion of exchange correlations, by solving the 3D
Gross-Pitaevksii equation for the condensate and using Thomas-Fermi
approximation for the degenerate Fermi gas. Our result shows that it only
plays a {\em minor} role compared with changing $a_{bf}$ from $-410a_0$ to $%
-330a_0$. Therefore, the inclusion of the possible effect of gravitational
sag may not be able to fully resolve the discrepancy mentioned in the paper.
\end{references}
\end{document}